\documentclass{ws-procs975x65}

\newcommand{\Mpc}{$h^{-1}$\thinspace Mpc}

\def\apj{ApJ}

\def\aa{A\&A}

\begin{document}


\title{Clusters and Superclusters in the Sloan and Las Campanas Surveys}

\author {J. Einasto}
\address{Tartu Observatory, EE-61602 T\~oravere, Estonia}

\maketitle

\abstracts{ We generate the 2-dimensional high-resolution density
field of galaxies of the Early Data Release of the Sloan Digital Sky
Survey and the Las Campanas Redshift Survey with a smoothing length
0.8~\Mpc\ to extract clusters and groups of galaxies, and a
low-resolution field with a smoothing length 10~\Mpc\ to extract
superclusters of galaxies.  We investigate properties of density field
clusters and superclusters and compare the properties of these
clusters and superclusters with those of Abell clusters, and
superclusters found on the basis of Abell clusters.  We found that
clusters in a high-density environment have a luminosity a factor of
$5 - 10$ higher than in a low-density environment. Clusters and
superclusters in the Northern slice of SDSS are much richer than those
in the Southern slice. }

\section{Introduction}

Clusters and groups of galaxies are the basic building blocks of the
Universe on cosmological scales.  The first catalogues of clusters of
galaxies (Abell 1958, Zwicky et al. 1961--68) were constructed by
visual inspection of the Palomar Observatory Sky Survey plates.
Cluster catalogues have been used to find superclusters of galaxies
(Oort~1983, Bahcall~1988, Einasto et al. 1994,~1997, 2001,
Basilakos~2003).

A major problem in the use of Abell clusters of galaxies is the
subjective character of the detection of clusters.  This problem can
be avoided if catalogues of galaxies of sufficient depth are
available.  Such catalogues are provided by the Las Campanas Redshift
Survey (LCRS) (Shectman et al. 1996) and by the Early Data Release of
the Sloan Digital Sky Survey (SDSS EDR) (Stoughton et al.  2002).
These surveys contain data on galaxies in relatively thin slices of
size of $1.5 \times 90$ degrees and of effective depth $z=0.2$.  Using
the LCRS and the SDSS EDR it is possible to calculate two-dimensional
density fields of galaxies and to detect groups and clusters of
galaxies as density peaks of the field. Similarly it is possible to
find superclusters of galaxies as large overdensity regions of the
density field. Here I shall report results of the determination of the
density fields of the LCRS and the SDSS EDR, and of the properties of
clusters and superclusters of galaxies found from the density field. A
more detailed description of our results is published elsewhere
(Einasto et al. 2003a, 2003b), see also the web-site of Tartu
Observatory (http://www.aai.ee).

\begin{table}[bt]
\tbl{Data on SDSS EDR and LCRS galaxies, clusters and superclusters}
{
         \label{Tab1}
         \begin{tabular}{rrrrrrrrr}
            \hline
            \noalign{\smallskip}
            Sample& DEC & RA & $\Delta$RA &
$N_{\rm gal}$ & $N_{\rm DF}$ & $N_{\rm LG}$ & $N_{\rm ACO}$ & $N_{\rm
            scl}$
\\ 

            \noalign{\smallskip}
            \hline
            \noalign{\smallskip}

SDSS.N& $0^{\circ}$& 190.2& 90.5 & 15209& 2868&&22&24 \\ 
SDSS.S& $0^{\circ}$&  23.2& 65.5 & 11882& 2287&&16&16\\ 
\\
LCRS&$-3^{\circ}$ & 191.4 & 81.0  &4065  & 1203  &289  & 18 & 19 \\
LCRS&$-6^{\circ}$ & 189.8 & 77.9  &2323  &  952  &147 & 13 &  17 \\
LCRS&$-12^{\circ}$& 191.4 & 81.1  &4482  & 1266  &276 & 11 &  15 \\
LCRS&$-39^{\circ}$&  12.1 &113.8  &3922  & 1285  &256 &  28 & 18 \\
LCRS&$-42^{\circ}$&  12.2 &112.5  &4158  & 1216  &265 &  19 & 14  \\
LCRS&$-45^{\circ}$&  12.3 &114.1  &3753  & 1182  &263 &  20 & 17  \\
\\
            \noalign{\smallskip}
            \hline
         \end{tabular}
}
   \end{table}

\section{Observational data}

The SDSS Early Data Release consists of two slices of about 2.5
degrees thick and $65-90$ degrees wide, centred on the celestial
equator, one in the Northern and the other in the Southern Galactic
hemisphere (Stoughton et al. 2002).  We obtained from the SDSS
Catalogue Archive Server the angular positions, Petrosian magnitudes,
and other available data for all the EDR galaxies. From this general
sample we extracted the Northern and Southern slice samples using the
following criteria: the redshift interval $1000 \leq cz \leq 60000$~km
s$^{-1}$, and the Petrosian $r^*$-magnitude interval $13.0 \leq r^*
\leq 17.7$.  The mean $DEC$, $RA$ and $RA$ intervals and the number of
galaxies extracted $N_{gal}$ are given in Table~\ref{Tab1}.  The
co-moving distances of galaxies were calculated using a cosmological
model with the following parameters: the matter density $\Omega
_m=0.3$, the dark energy density $\Omega _\Lambda =0.7$, the total
density $\Omega _0=\Omega _m+\Omega _\Lambda =1.0$, all in units of
the critical cosmological density (see Peacock 1999). With these
parameters the limiting redshift $z_{lim}=0.2$ corresponds to the
co-moving distance $r_{lim}=571$~\Mpc.  In calculating absolute
magnitudes we used K-corrections and the correction for absorption in
the Milky Way.

The LCRS (Shectman et al. 1996) is an optically selected galaxy
redshift survey that extends to a redshift of 0.2 and covers six
slices of width of 1.5 degrees, 3 slices are located in the Northern
Galactic cap, and 3 in the Southern cap.  The spectroscopy of the
survey was carried out via a 50 or a 112 fibre multi-object
spectrograph; therefore the selection criteria vary from field to
field. The nominal apparent magnitude limits for the 50 fibre fields
are $m_1=16.0 \le R \le m_2=17.3$, and for the 112 fibre fields
$m_1=15 \le R \le m_2=17.7$.  The data on the LCRS slices are given in
Table~\ref{Tab1}.

We also use the sample of rich clusters of galaxies by Abell (1958)
and Abell et al. (1989) (hereafter Abell clusters), compiled by
Andernach \& Tago (1998), with redshifts up to $z=0.13$.  This sample
was described in detail by Einasto et al. (2001), where an updated
supercluster catalogue of Abell clusters was presented.  Superclusters
were identified as clusters of Abell clusters using a
friend-of-friends algorithm with a neighbourhood radius of 24~\Mpc.
For the LCRS we use also the catalogue of loose groups with at least 3
observed galaxies by Tucker et al.~(2000).

\section{Density field clusters}

\subsection{The high-resolution density field of the SDSS}

We shall use the density field to find density enhancements -- density
field clusters, or shortly DF-clusters.  The procedure consists of the
following steps: (1) calculation of the distance, the absolute
magnitude, and a weight factor for each galaxy of the sample; (2)
calculation of rectangular coordinates of galaxies; (3) smoothing of
the density field using an appropriate kernel and a smoothing length,
and (4) finding clusters (density enhancements) in the field.  When
calculating the density field we regard every galaxy as a visible
member of a density enhancement (group or cluster) within the visible
range of absolute magnitudes, $M_1$ and $M_2$, corresponding to the
observational window of apparent magnitudes at the distance of the
galaxy. This assumption is based on observations of nearby galaxies,
which indicate that practically all giant galaxies are surrounded by
dwarf companions, like our own Galaxy or the M31.

The expected number and luminosities of faint galaxies outside the
visibility window is calculated using the Schechter~(1976) luminosity
function with parameters given by Einasto et al. (2003a, 2003b).  When
calculating of the total luminosity of the group, we have encountered
two selection effects.  When using Schechter parameters found with
conventional methods we multiply the luminosity of the observed galaxy
by a weight which takes statistically into account all galaxies
outside the visibility window.  As we shall see below, on large
distances faint groups and clusters of galaxies are completely missed
since they contain no objects inside the visibility window.  To take
these galaxies into account their luminosity is added to the
luminosity of visible groups and clusters.  As a result the visible
groups and clusters become too bright, but the summed luminosity of
all groups and clusters at a given distance is correct.  Thus in this
case we get correct properties of superclusters of galaxies, which are
defined as large overdensity regions.  To get correct properties for
individual groups and clusters we have used another set of Schechter
parameters which do not add the luminosity of completely invisible
groups to the luminosity of visible ones.  With this set of parameters
we have calculated all properties of DF-clusters, their luminosity
distribution etc.

\begin{figure*}[ht]
\centering
\resizebox{.65\columnwidth}{!}{\includegraphics*{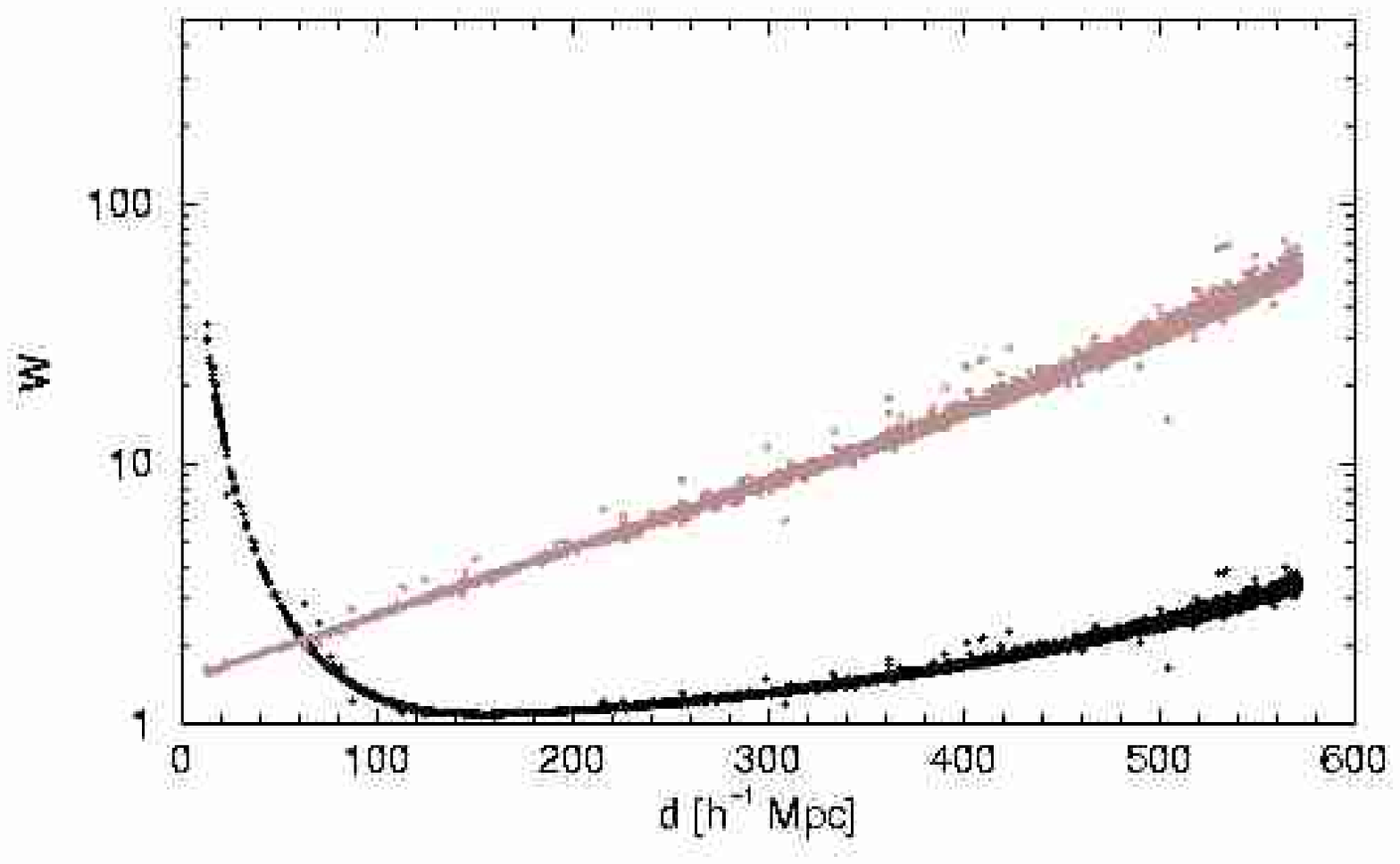}}\\
\resizebox{.65\columnwidth}{!}{\includegraphics*{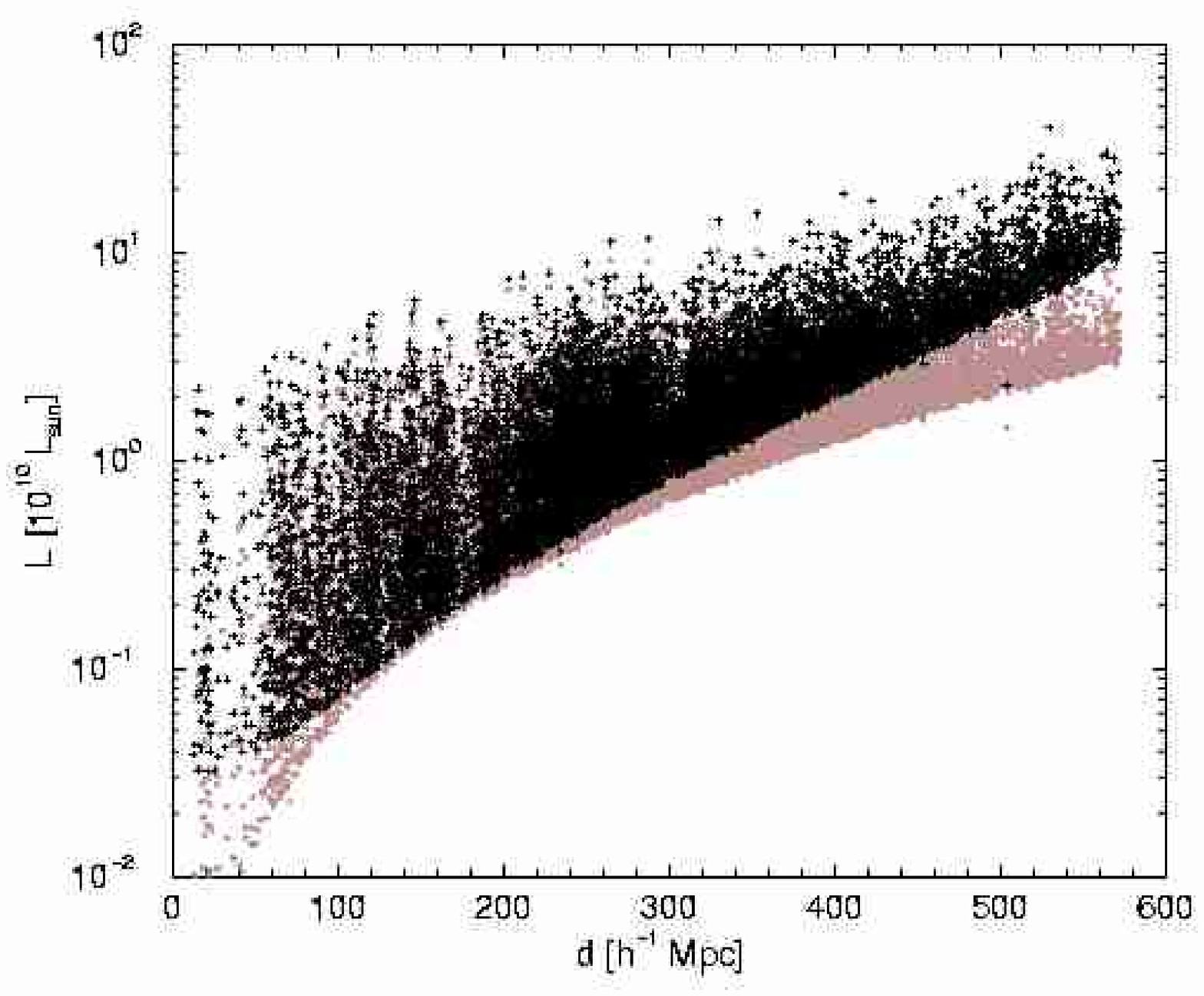}}
\caption{\label{fig:1} The upper panel shows the weights as a function of
  distance.  Grey symbols indicate the number-density weights, black
  symbols -- the luminous-density weights.  The lower panel plots
  luminosities of galaxies as a function of distance.  Grey circles
  mark luminosities of observed galaxies, black symbols mark total
  luminosities, corrected for the unobservable part of the luminosity
  range.  Here we use the luminosity function parameters of set 1, which
  yield better DF-clusters. }
\end{figure*}

As an example, we plot in Fig.~\ref{fig:1} number-density and
luminous-density weights for the luminosity function which yields
correct properties for individual DF-clusters (for the Northern
slice). Also we show the luminosities of galaxies $L_{obs}$, and the
expected total luminosities $L_{tot}$.  The luminosities are expressed
in the units of $10^{10}$ Solar luminosities.  We see that the
number-density weights $W_N$ rise monotonically with the increasing
distance from the observer, whereas the luminous-density weights $W_L$
rise also toward very small distances. This is due to the influence of
bright galaxies outside the observational window, which are not
numerous, but are very luminous.

To find the luminous-density field we calculated rectangular
equatorial coordinates for all galaxies.  The coordinates were rotated
to obtain a situation where the new $y$-axis was oriented toward the
average right ascension of each slice, and was rotated around the
$x$-axis so as to force the average $z$-coordinate to be zero in order
to minimise the projection effects.  The SDSS slices are located
around the celestial equator, thus the last rotation angle is in this
case zero.

\begin{figure*}[ht]
\centering
\resizebox{.85\columnwidth}{!}{\includegraphics*{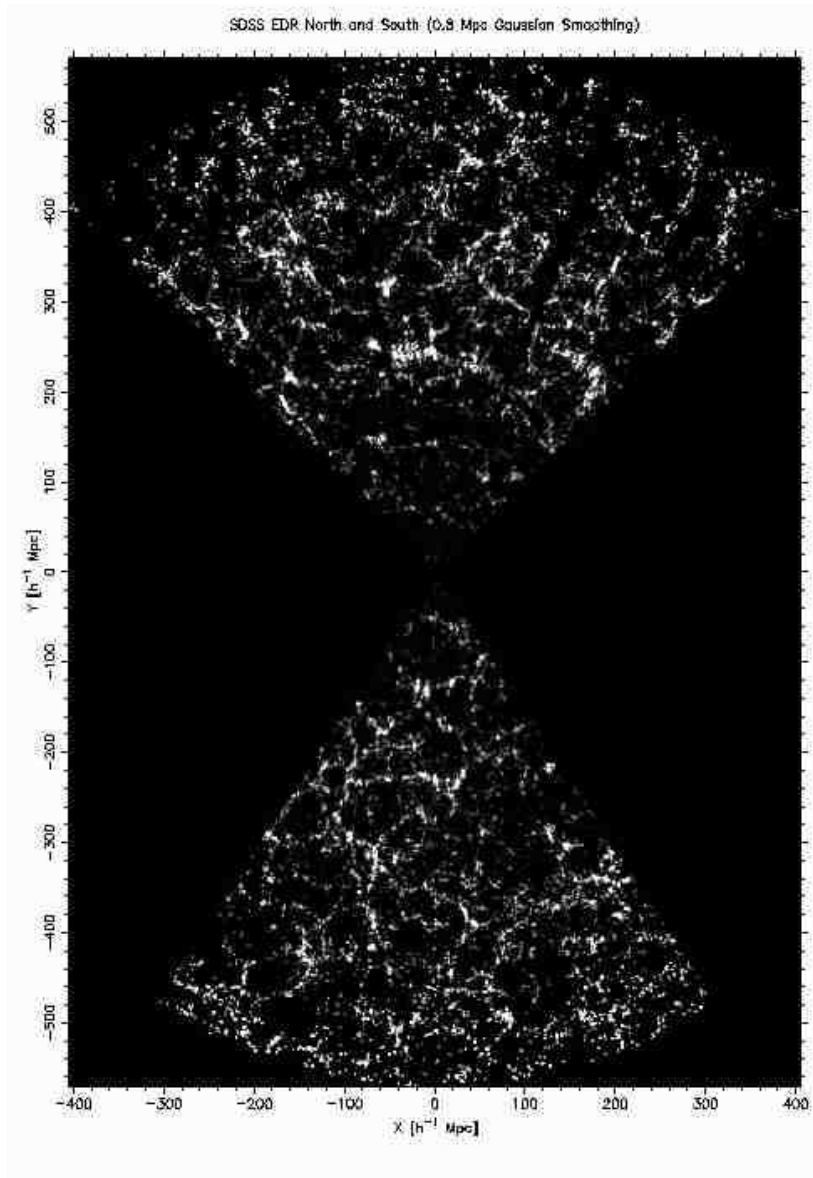}}
\caption{The density field of the SDSS EDR slices, smoothed with a
  Gaussian of $\sigma = 0.8$~\Mpc\ dispersion.  } 
\label{fig:2}
\end{figure*}

To calculate the density fields, we formed a grid of a cell size of
1~\Mpc, and used only the $x,y-$coordinates.  Ignoring the
$z-$coordinate is equivalent to integrating over this coordinate,
i.e. we find actually the projected density of the wedge.  Finally the
density field was smoothed using the Gaussian smoothing.  To find
compact overdensity objects -- the DF-clusters -- we used a smoothing
length of 0.8~\Mpc, this field is shown in Fig.~\ref{fig:2}.  Since
the thickness of the wedge increases with the distance from the
observer, the number of DF-clusters per unit surface area also
increases.

\subsection{DF-clusters}

DF-clusters of galaxies were identified as peaks of the
high-resolution density field.  The total luminosity of a DF-cluster
was derived by adding the luminosity densities of $5 \times 5$ cells
centred at the cell with the highest luminosity density.  This
corresponds to the dispersion 0.8~\Mpc\ used in smoothing.  The
luminosities were calculated in Solar luminosity units.  We also
calculated the density in units of the mean density, this relative
density is used in Fig.~\ref{fig:2} for plotting.  A cluster was added
to the list of DF-clusters if its luminosity $L$ exceeded the
threshold value $L_0 = 0.4 ~10^{10}~ L_{\odot}$, and its distance
was in the interval $100 \leq d \leq 550$~\Mpc.  The total number of
DF-clusters found is given in Table~\ref{Tab1}.

\begin{figure*}[ht]
\centering
\resizebox{.65\columnwidth}{!}{\includegraphics*{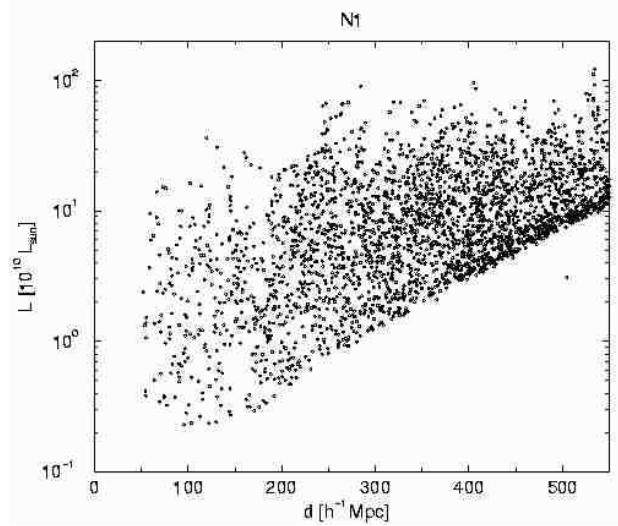}}
\caption{Luminosities of DF-clusters as a function of distance  
for the Northern slice, for the  parameter set 1. } 
\label{fig:3}
\end{figure*}

Now we look at some properties of DF-clusters.  Fig.~\ref{fig:3} shows
the luminosities of DF-clusters as a function of the distance from the
observer, $d$.  The lowest luminosity clusters are seen only at
distances $d \le 150$~\Mpc.  There exists a well-defined lower limit
of cluster luminosities at larger distances, the limit being linear in
the $\log L - d$ plot.  Such behaviour is expected, as at large
distances an increasing fraction of clusters does not contain any
galaxies bright enough to fall into the observational window of
absolute magnitudes, $M_1 \dots M_2$.

\subsection{The luminosity function of DF-clusters}

\begin{figure}[ht]
\centering
\resizebox{.65\columnwidth}{!}{\includegraphics*{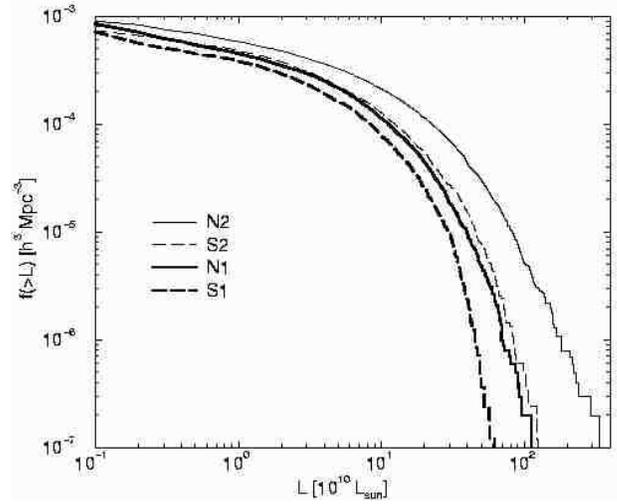}}
\caption{Integrated luminosity functions of DF-clusters for the
parameter sets 1 and 2.  }
\label{fig:4}
\end{figure}

Now we calculate the integrated luminosity function of DF-clusters,
i.e. the number of DF-clusters per unit volume exceeding a given
luminosity $L$.  Fig.~\ref{fig:3} shows that only very bright
DF-clusters are observable over the whole depth of our samples.  The
estimated total number of fainter clusters can be found by multiplying
the observed number of clusters by a weighting factor
$(d_{lim}/d_L)^3$, where $d_{lim} = 550$~\Mpc\ is the limiting
distance used in compiling the DF-cluster sample, and $d_L$ is the
largest distance where clusters of luminosity $L$ are seen. The
luminosity functions of DF-clusters for both the SDSS samples are
shown in Fig.~\ref{fig:4}.

We see that the luminosity functions span over 3 orders in luminosity
and almost 4 orders in spatial density.  The bright DF-clusters are
the density field equivalents of rich clusters of the Abell class,
fainter DF-clusters correspond to groups of galaxies. The broad
coverage of luminosities shows that the SDSS sample is well suited for
determining the luminosity function from observations.  The number of
bright clusters is much larger for the parameter set 2.  As discussed
above, this is due to the overcorrection of cluster luminosities --
particularly at large distances -- to compensate for non-detected
faint clusters. Thus we believe that the function for the parameter
set 1 corresponds better to reality.  Our results also show that the
Northern SDSS sample has a larger number of luminous DF-clusters than
does the Southern sample.

\subsection{Environmental effects in the distribution of DF-clusters}

\begin{figure*}[ht]
\centering
\resizebox{0.48\textwidth}{!}{\includegraphics*{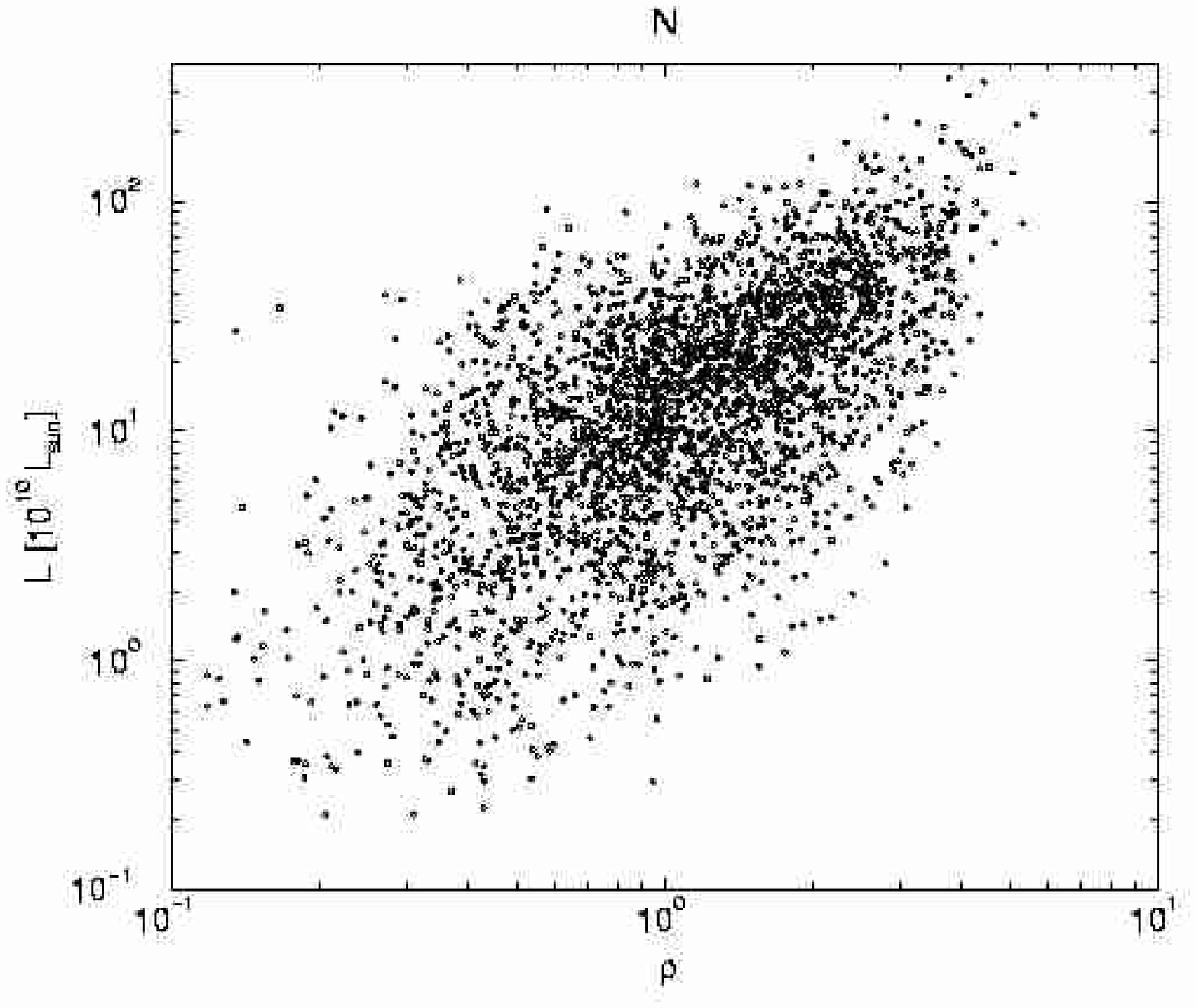}}\hspace{2mm}
\resizebox{0.48\textwidth}{!}{\includegraphics*{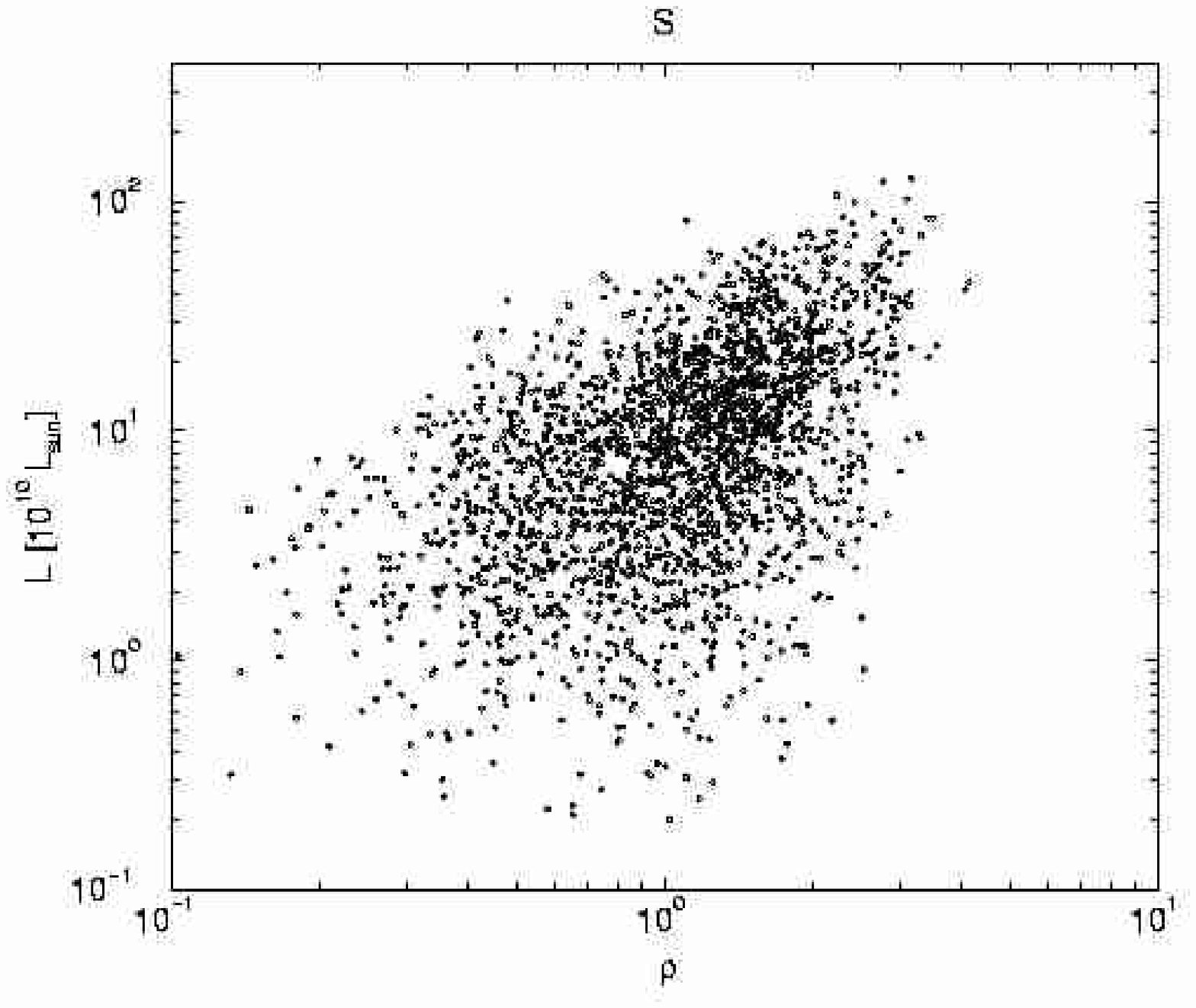}}\hspace{2mm}
\caption{The luminosities of DF-clusters as a function of the global
relative density $\varrho_0$.  The left panel shows the Northern
slice, the right panel the Southern slice. }
\label{fig:5}
\end{figure*}

We can use the density found with the 10~\Mpc\ smoothing as an
environmental parameter to describe the global density in the
supercluster environment of clusters.  We calculated this global
density $\varrho_0$ (in units of the mean density of the
low-resolution density field) for all the DF-clusters.
Fig.~\ref{fig:5} shows the luminosity of DF-clusters as a function of
the global density $\varrho_0$.  There is a clear correlation between
the luminosity of DF-clusters and their environmental density.
Luminous clusters are predominantly located in high-density regions,
and low-luminosity clusters in low-density regions.  This tendency is
seen also visually in the colour version of Fig.~\ref{fig:2} (see
Einasto et al. 2003a and the web-site of Tartu Observatory).  Here
densities are colour-coded, and we see that small clusters in voids
have blue colours which indicate medium and small densities, whereas
rich clusters having red colours populate dominantly the central
high-density regions of superclusters.

\begin{figure*}[ht]
\centering
\resizebox{.50\columnwidth}{!}{\includegraphics*{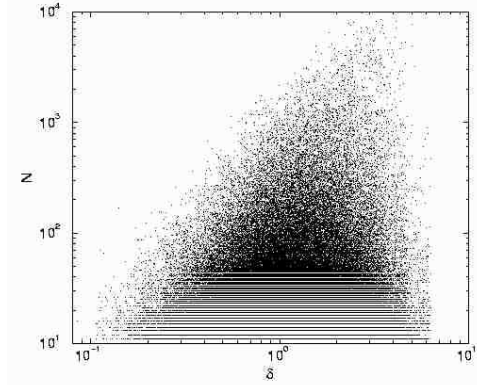}}
\caption{The masses of clusters in numerical simulation as a function
of the global relative density $\varrho_0$.  }
\label{fig:6}
\end{figure*}

For comparison we show in Fig.~\ref{fig:6} the masses of clusters in
numerical simulations as a function of the global relative density.
The numerical model was calculated for a box of size 200~\Mpc, using
the cosmological parameters: $\Omega_m=0.3$, $\Omega_{\Lambda} = 0.7$.
The masses of clusters are characterised by the number of particles in
clusters, $N$.  We see that in high-density environments most massive
clusters are more than a hundred times more massive than in
low-density environments.  The contrast is larger than that found for
the observed DF-clusters.  Probably this is due to a higher mass
resolution in numerical simulations.

Fig.~\ref{fig:5} shows also that the Northern sample has much more
luminous clusters than the Southern sample.  This result confirms the
presence of a difference between the structure of Northern and
Southern samples.

\begin{figure*}[ht]
\centering
\resizebox{0.48\textwidth}{!}{\includegraphics*{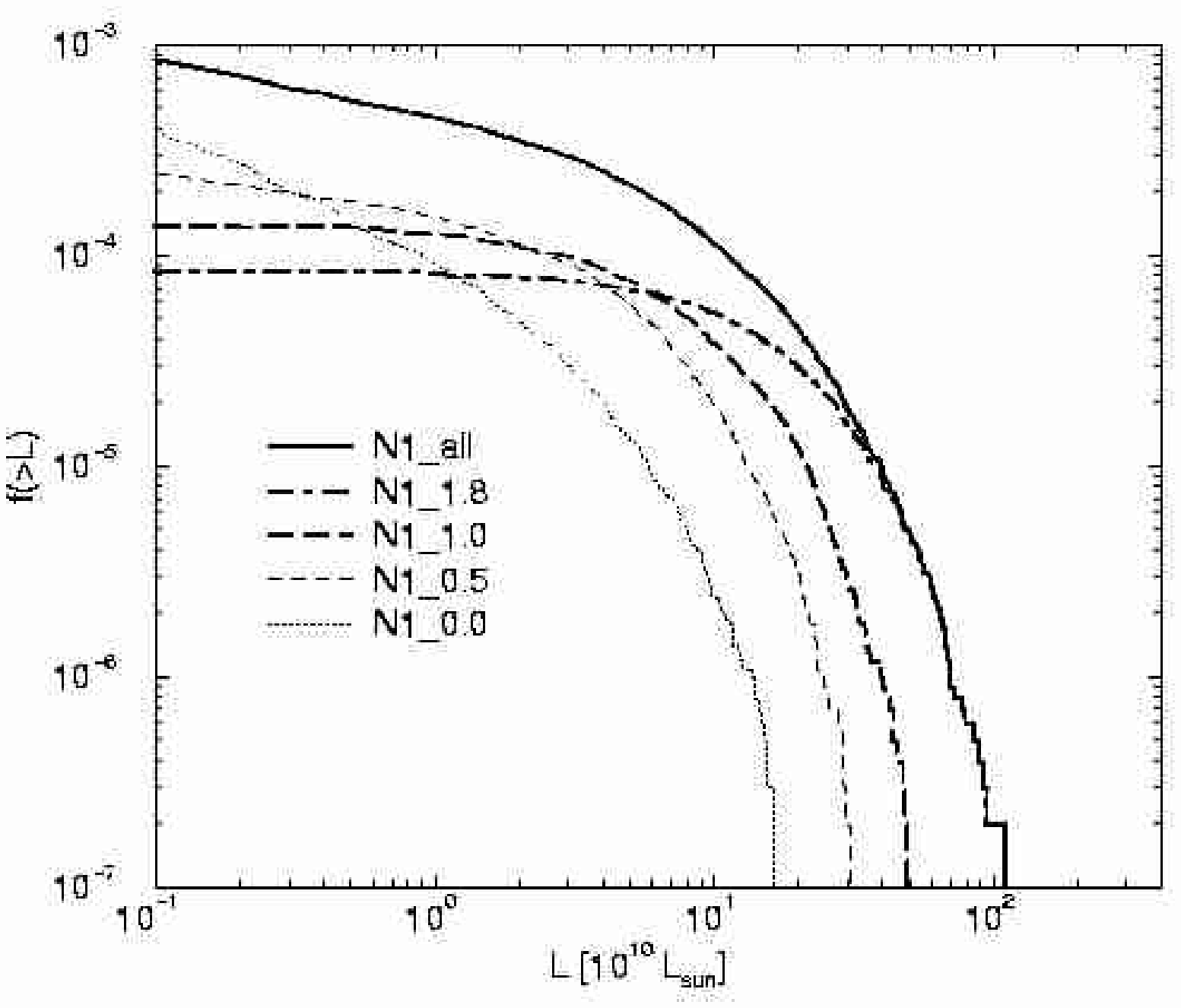}}\hspace{2mm}
\resizebox{0.48\textwidth}{!}{\includegraphics*{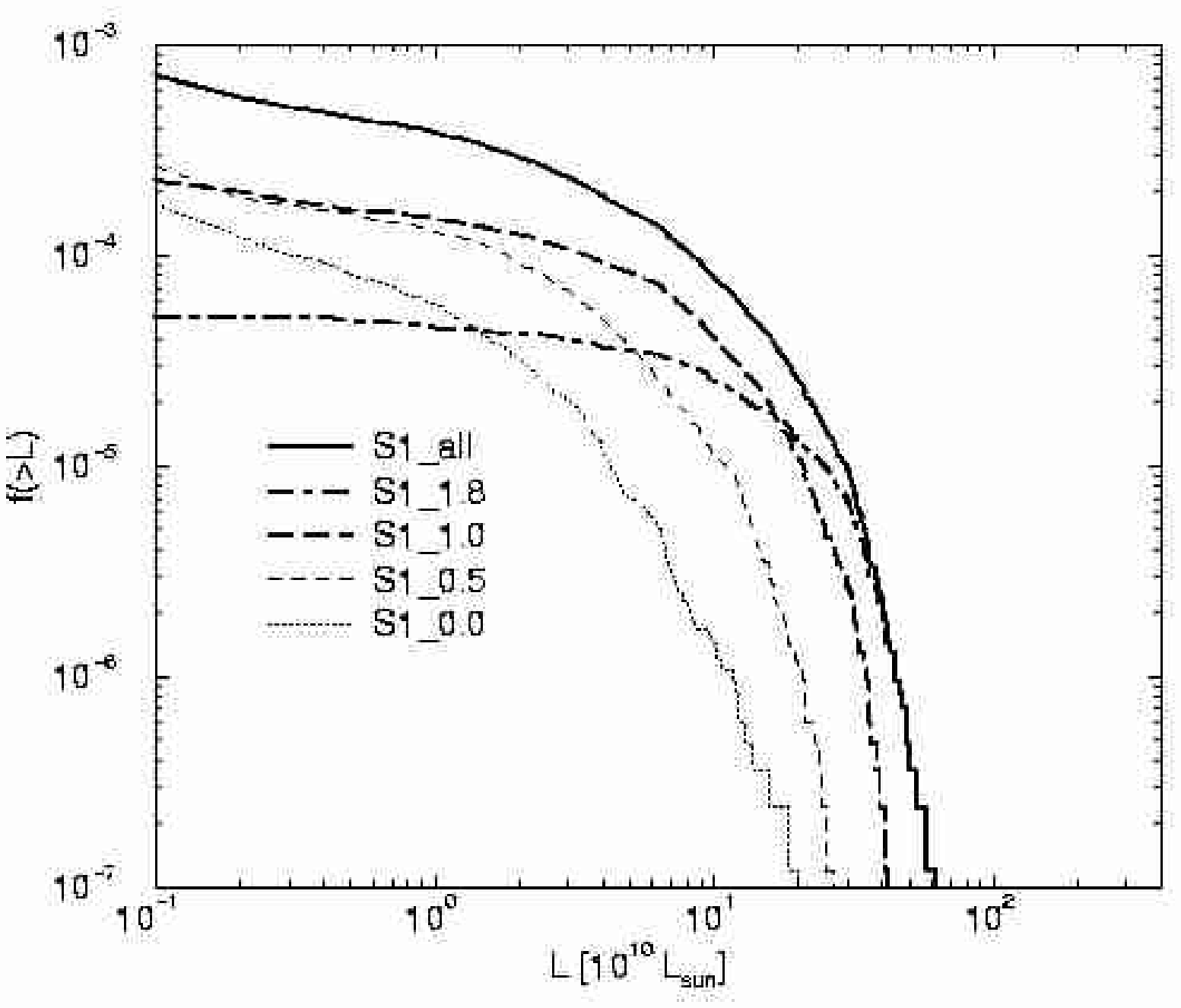}}\hspace{2mm}
\caption{The integrated luminosity functions of the DF-clusters for various
 global relative density intervals $\varrho_0$.  The left panel shows the
 Northern slice, the right panel -- the Southern slice. }
\label{fig:7}
\end{figure*}

Fig.~\ref{fig:7} shows the cluster luminosity functions of the
Northern and Southern slices calculated separately for 4 global
density intervals, $0 < \varrho_0 \leq 0.5$, $0.5 < \varrho_0 \leq
1.0$, $1.0 < \varrho_0 \leq 1.8$, and $1.8 < \varrho_0 \leq 10$,
labelled in Fig.~\ref{fig:7} as N0.0, N0.5, N1.0 and N1.8, for the
Northern subsamples, and S0.0, S0.5, S1.0, S1.8 for the Southern ones.
The functions were calculated for the luminosity function parameter
set 1 (which yields an almost constant highest luminosity for the
DF-clusters for various distances from the observer).  We see that the
luminosity functions depend very strongly on the global environment:
the luminosity of the most luminous clusters in subsamples differs by
a factor of $5 \pm 2$ (for a finer global density division the most
luminous DF-clusters differ in luminosity even up to 10 times in
extreme density regions).  This difference is not statistical, due to
different numbers of clusters, as these numbers are all of the same
order.

\section{Density field superclusters}

\subsection{The supercluster catalogue}

Superclusters have been traditionally defined either as clusters of
clusters of galaxies (Oort 1983, Bahcall 1988), using Abell clusters
(Einasto et al. 1994, 1997, 2001), or as large and rich systems of
galaxies (de Vaucouleurs 1953). In the present analysis we use the
low-resolution density 
field of galaxies to find large overdensity regions which we call
density field superclusters or for short DF-superclusters.

\begin{figure*}[ht]
\centering
\resizebox{.85\columnwidth}{!}{\includegraphics*{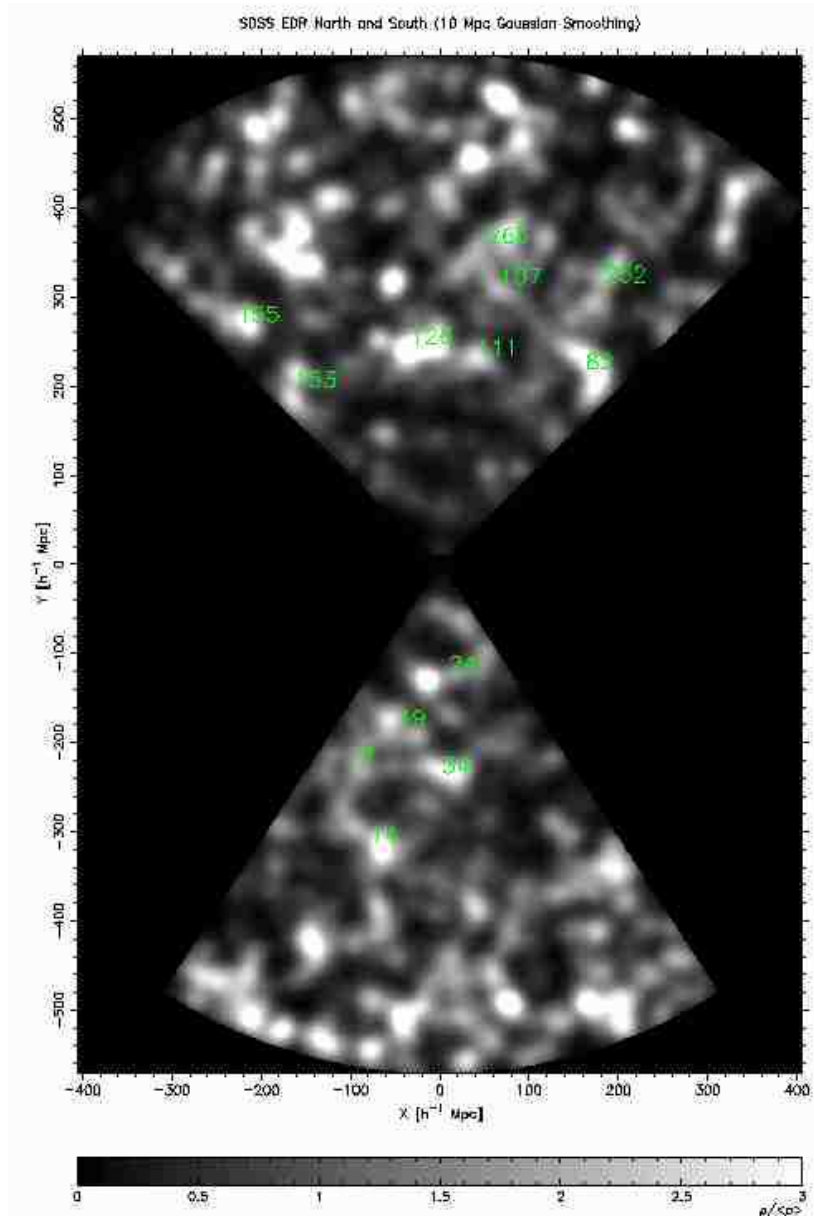}}
\caption{The density field of the SDSS EDR slices, smoothed with
$\sigma = 10$~\Mpc.  The numbers denote Abell superclusters
  according to Einasto et al. (2001).} 
\label{fig:8}
\end{figure*}

\begin{figure*}
\centering
\resizebox{0.45\textwidth}{!}{\includegraphics*{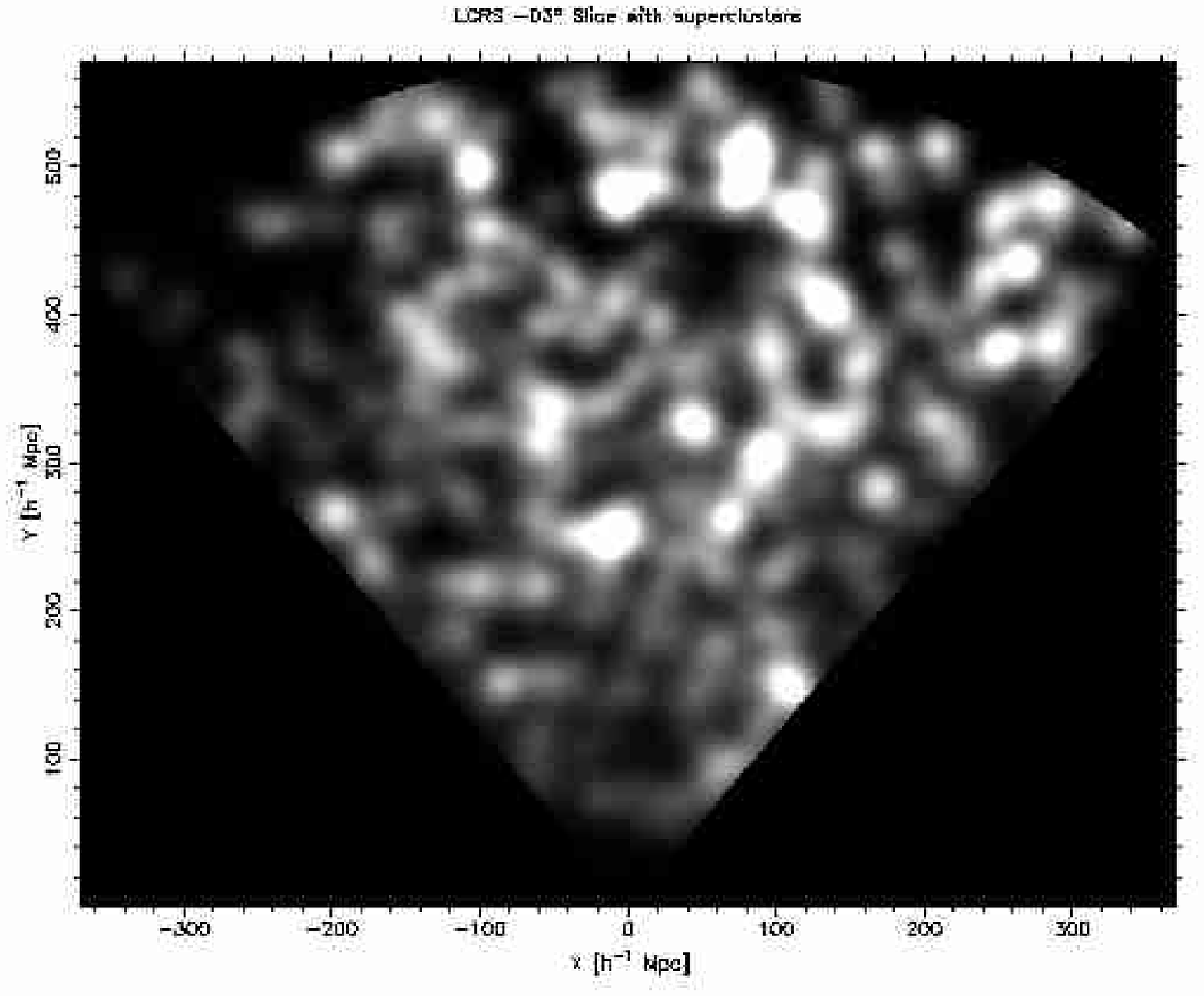}}\hspace{2mm}
\resizebox{0.45\textwidth}{!}{\includegraphics*{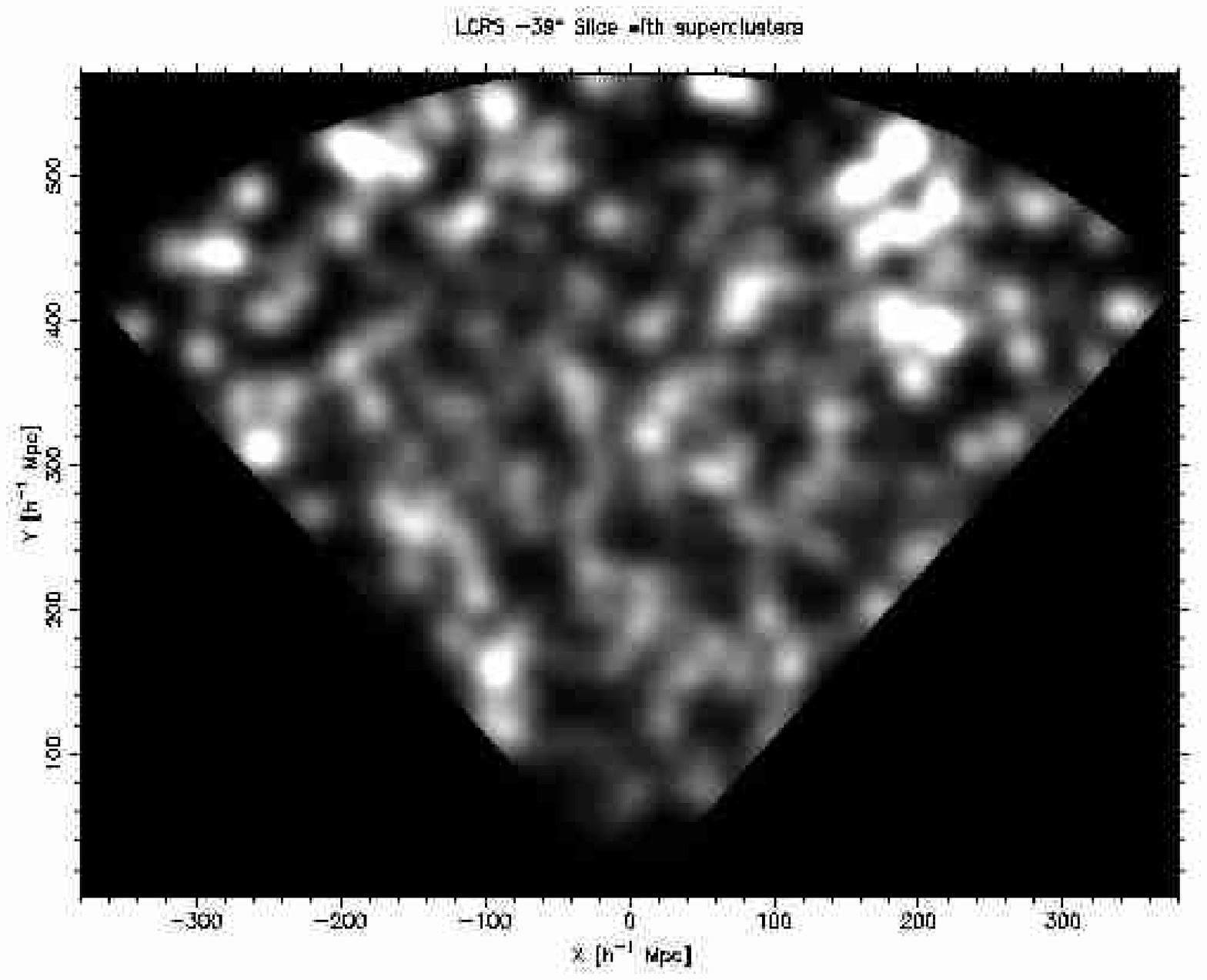}}\hspace{2mm}\\
\resizebox{0.45\textwidth}{!}{\includegraphics*{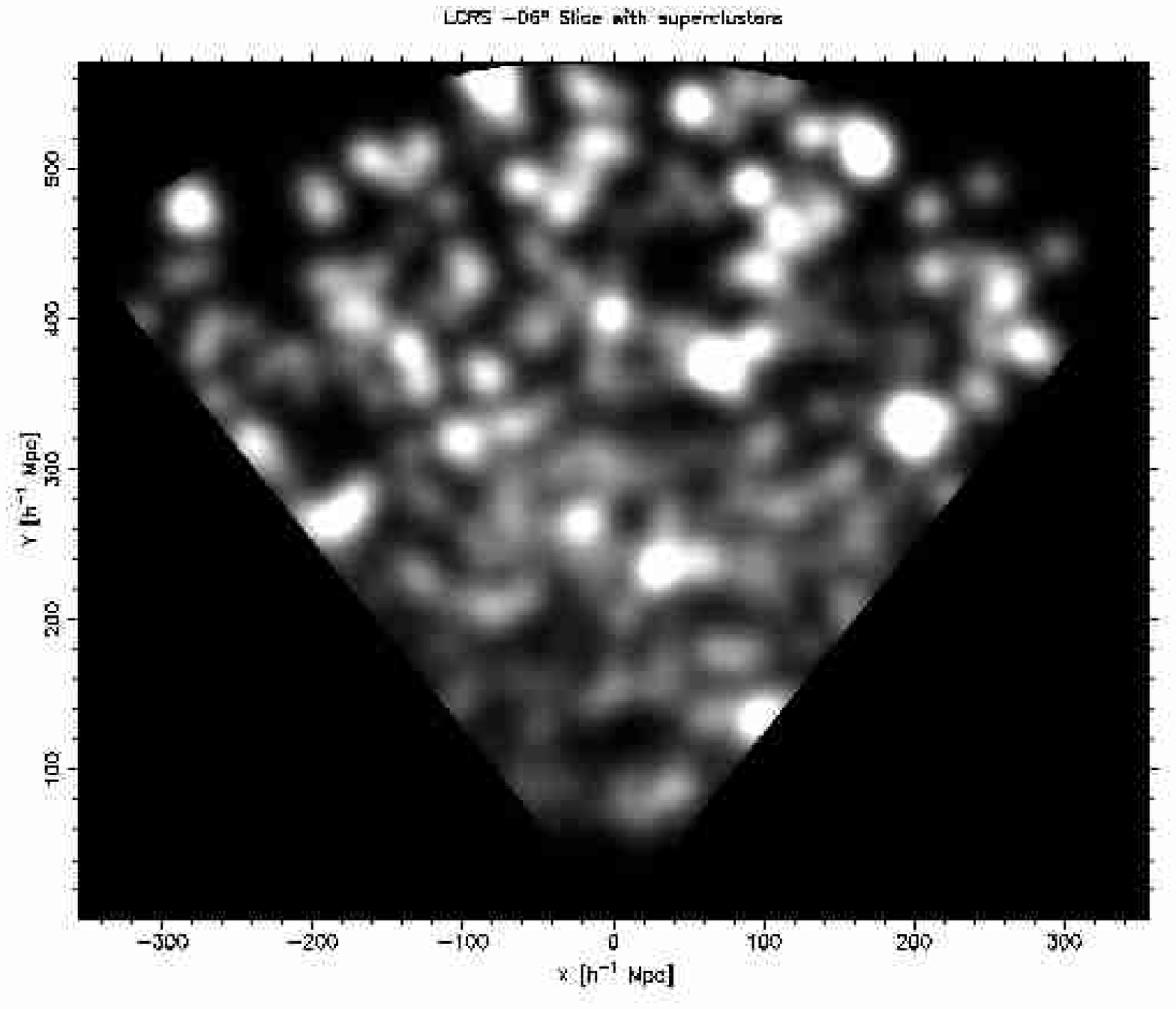}}\hspace{2mm}
\resizebox{0.45\textwidth}{!}{\includegraphics*{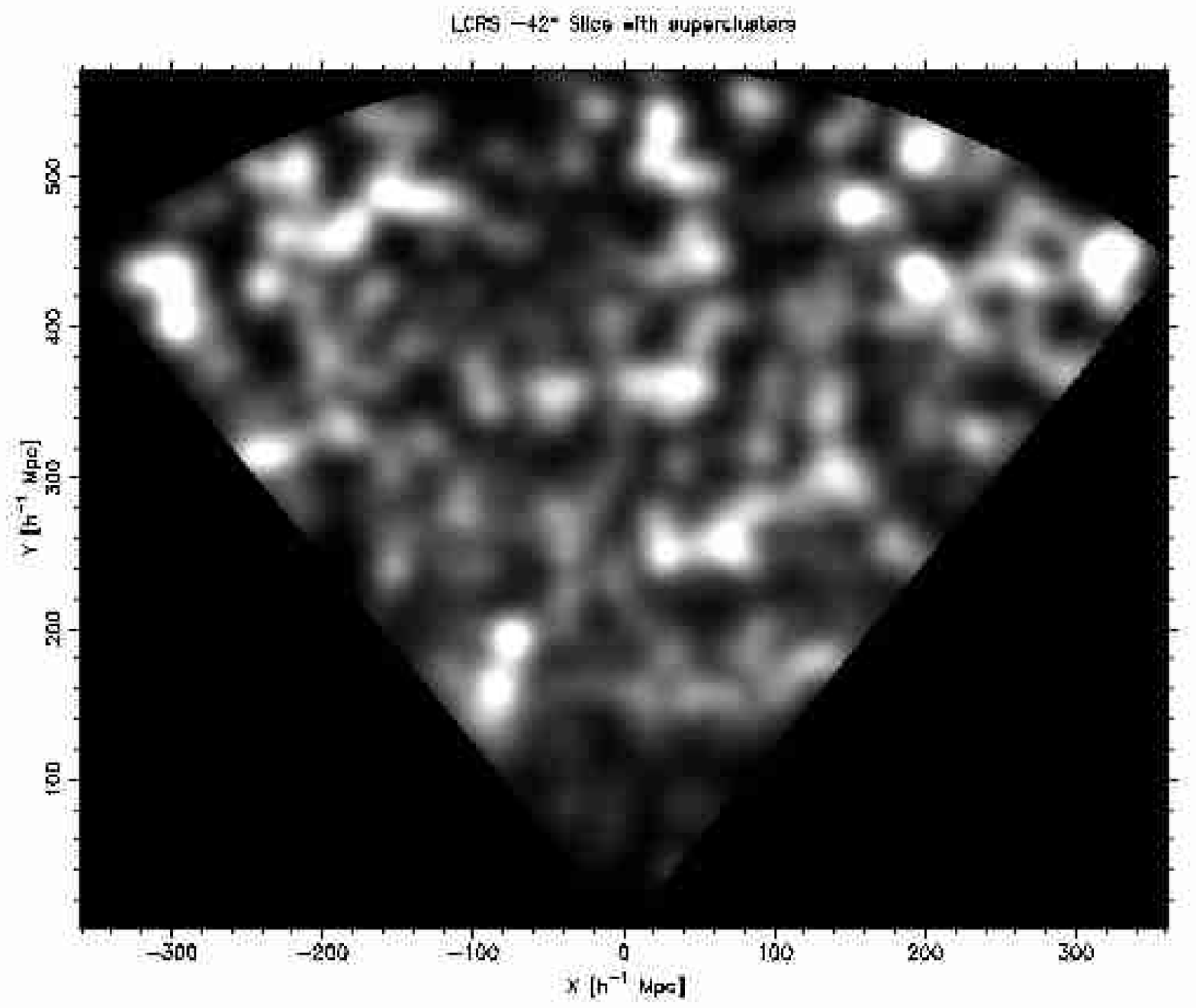}}\hspace{2mm}\\
\resizebox{0.45\textwidth}{!}{\includegraphics*{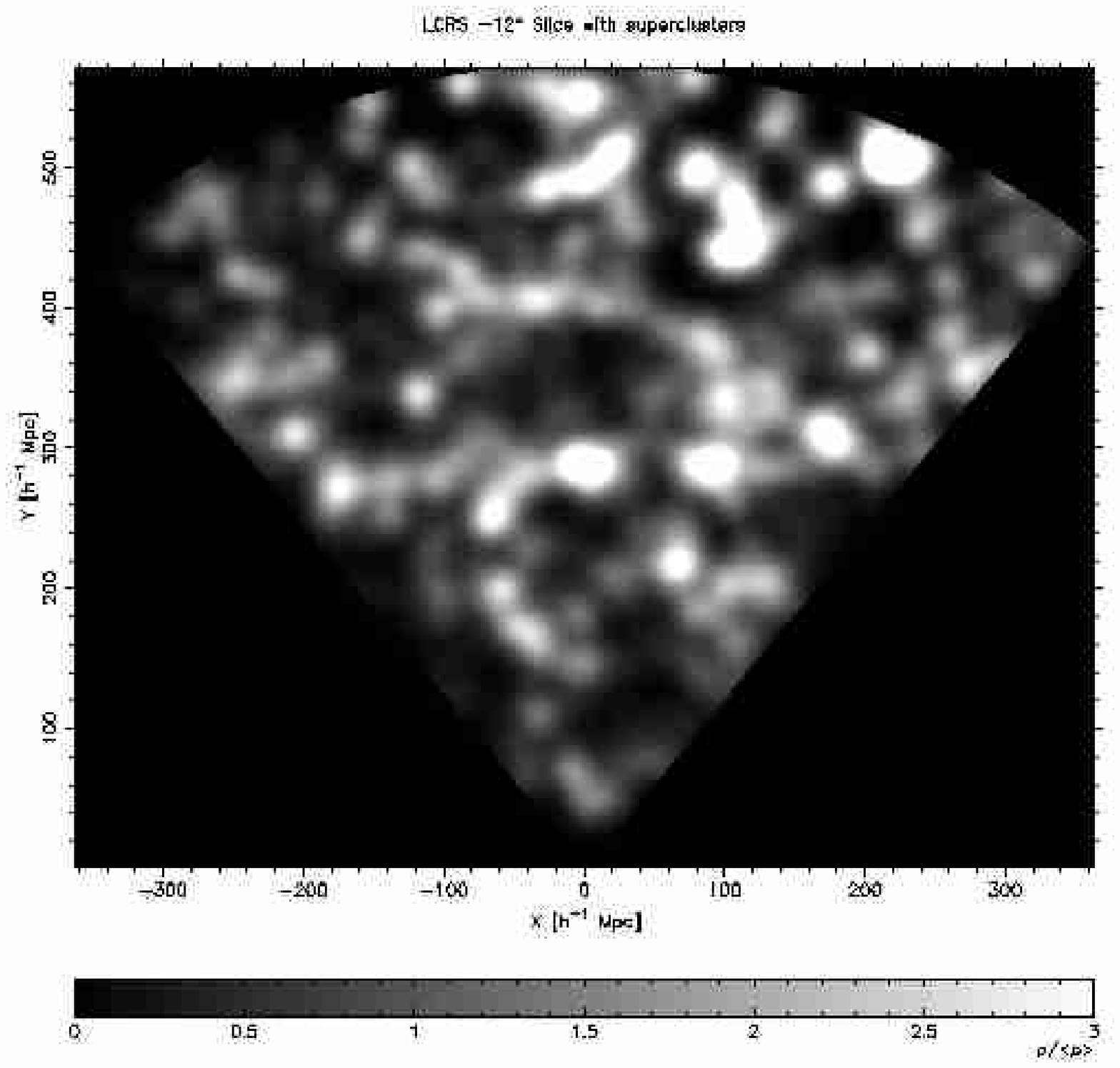}}\hspace{2mm}
\resizebox{0.45\textwidth}{!}{\includegraphics*{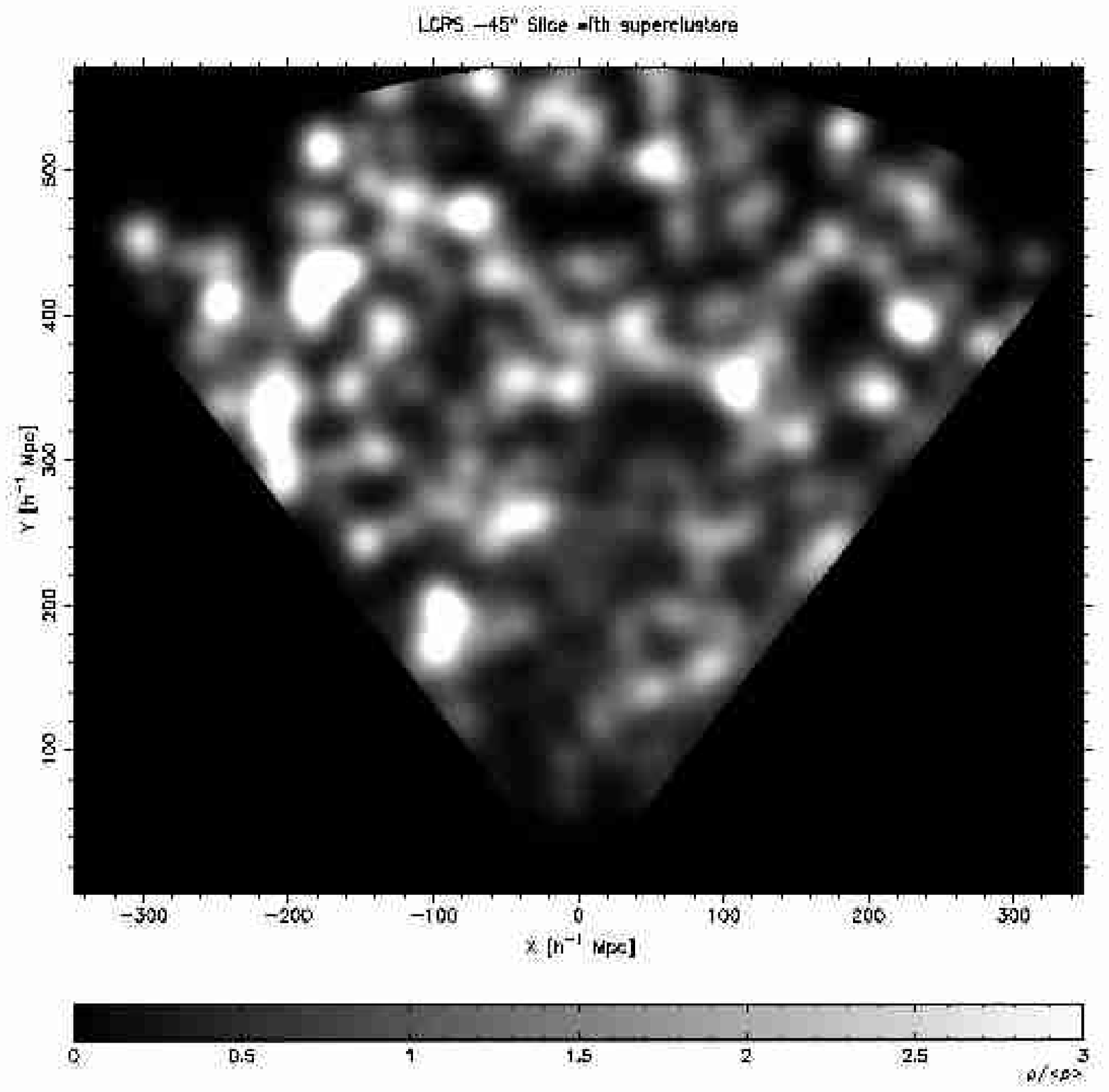}}\hspace{2mm}\\
\caption{The density field of the LCRS slices smoothed with a
$\sigma = 10$~\Mpc\ Gaussian filter. The densities are reduced to a sheet
  of constant thickness.}   
\label{fig:9}
\end{figure*}

The compilation of the supercluster catalogue consists of three steps:
(1) calculating the density field, 2) finding the overdensity regions, and
3) determining the properties of the resulting superclusters.  The
density field was calculated as described above with one difference --
in order to reduce the wedge-like volume of slices to a sheet of
uniform thickness we divided densities by the thickness of the slice
at each particular distance.  In this way the surface density of the
field is on average independent of distance, and we can use a distance
independent search for overdensity regions.  The reduced density field
for the Northern and Southern SDSS EDR slices is shown in
Fig.~\ref{fig:8}.  Fig.~\ref{fig:9}  shows the
low-resolution density field of the six sheets of the Las Campanas
Redshift Survey.  The field was calculated similarly to the
low-density field of the Sloan Survey.

\begin{figure}[ht]
\centering
\resizebox{.65\columnwidth}{!}{\includegraphics*{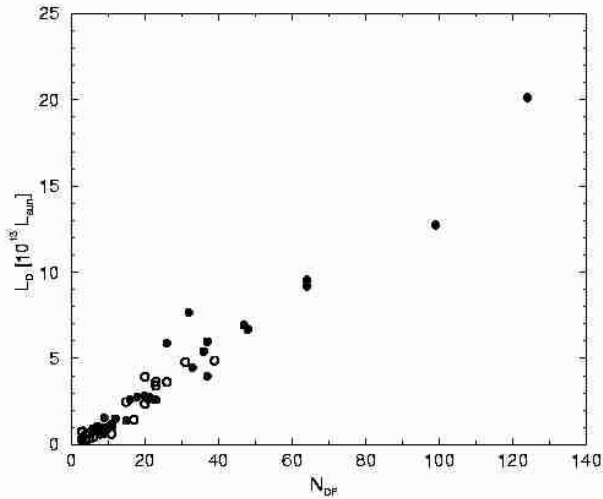}}
\caption{The total luminosities of the DF-superclusters versus the number
of DF-clusters in a supercluster (supercluster richness). Filled
circles: DF-superclusters which belong to the superclusters of Abell
clusters, open circles: DF-superclusters that do not belong to the
Abell superclusters.  }
\label{fig:10}
\end{figure}

Next we searched for connected high-density regions. To do so we need
to fix the threshold density, $\varrho_0$, which divides the high- and
low-density regions.  This threshold density plays the same role as
the neighbourhood radius used in the friends-of-friends (FoF) method
to find clusters in galaxy samples and to find superclusters in
cluster samples.  To make a proper choice of the threshold density we
calculated the number of superclusters, the area of the largest
supercluster, and the maximum diameter of the largest supercluster as
a function of the threshold density $\varrho_0$.  This test shows that
a proper value of the threshold density is $\varrho_0 = 1.8$ (we use
relative densities here). In this case most superclusters are well
separated, only one large supercluster consists of several density
enhancements, which are separated at $\varrho_0= 2.1$.  Superclusters
were found for the distance interval $100 \le d \le 550$~\Mpc, with
areas greater than 100~(\Mpc)$^2$.  The number of superclusters in the
SDSS and LCRS slices is shown in Table~\ref{Tab1}.

We also calculated the observed total luminosity of a supercluster
$L_{obs}$ (the sum of the observed luminosities of the DF-clusters
located within the boundaries of a supercluster), and the total
luminosity of the supercluster $L_{tot}$.  The total luminosity was
calculated from the observed luminosity under the assumption that the
vertical diameter of the supercluster is identical to the diameter in
the plane of the slice.  Using the spatial distribution of the
DF-clusters we estimated also the morphological type of the
supercluster.

\subsection{The fine structure of superclusters and voids}

\begin{figure*}[ht]
\centering
\resizebox{.95\columnwidth}{!}{\includegraphics*{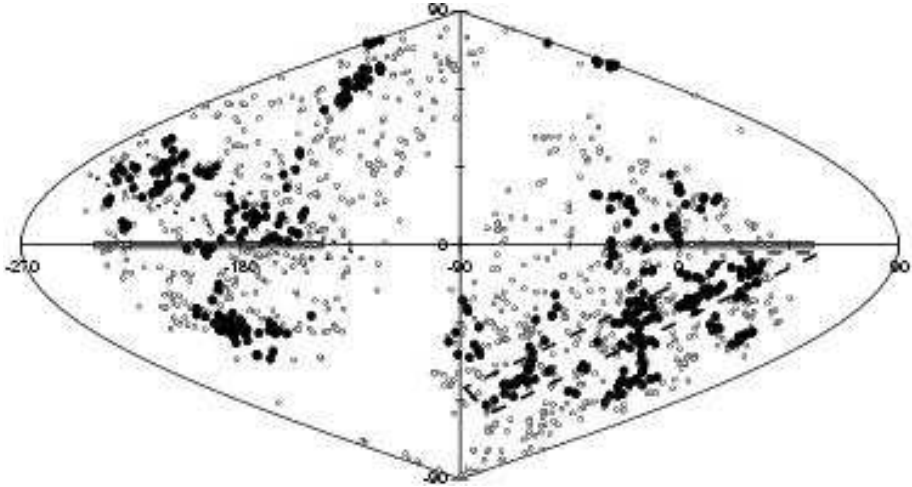}}
\caption{A view of Abell clusters in equatorial coordinates. Filled
circles show the Abell clusters located in superclusters of richness 8
and more members, open circles mark the Abell clusters in less rich
superclusters.  The strips near the celestial equator mark the SDSS
slices, the Northern slice is at the left side.  The areas surrounded
by dashed lines indicate regions where rich superclusters of the
Dominant Supercluster Plane are located.  The Galactic zone of
avoidance is a S-shaped curve in equatorial coordinates. }
\label{fig:11}
\end{figure*}

The distribution of DF-clusters within DF-superclusters yields
information on the internal structure of superclusters.  A close
inspection of Figs.~\ref{fig:2} and \ref{fig:8} shows that
superclusters have different internal structures: clusters may form a
single filament, a branching system of filaments, or a more or less
diffuse cloud of clusters.  Most rich superclusters have a
multi-filamentary morphology, poor superclusters consist usually of one
cluster filament, or have a compact morphology without a clear
filamentary system.  Most superclusters are surrounded by faint
systems of galaxies, either in the form of filaments or of a diffuse
cloud of clusters.

In the SDSS and LCRS samples about $15 - 25$\%\ of all DF-clusters are
located in superclusters.  Most DF-clusters outside superclusters also
form filaments.  Thus we come to the conclusion that there is no major
difference in the shape of cluster systems in supercluster and in
non-supercluster environments. This similarity of the structure within
and outside superclusters is partly due to our formal procedure of
defining superclusters; actually there is a continuous sequence of
structures from single filaments to multiple filaments and
superclusters.  The difference is mainly in the luminosity of
clusters.  This observation can be interpreted as follows: clusters
and cluster filaments in various environments are formed by similar
density perturbations.  Small-scale perturbations are modulated by
large-scale perturbations which make clusters and their filaments
richer in superclusters and poorer in large voids between
superclusters (see also Frisch et al. 1995).

Fig.~\ref{fig:10} shows the total luminosities of superclusters versus
their richnesses (the number of DF-clusters in a supercluster).  This
figure shows that those DF-superclusters that are also the Abell
superclusters are more luminous and richer than the non-Abell
DF-superclusters.  Einasto et al. (2003b, 2003c, 2003d) showed, using
the data on the Las Campanas loose groups (Tucker et al. 2000), that
the loose groups in superclusters of Abell clusters are richer, more
luminous, and more massive than the loose groups in systems that do
not belong to Abell superclusters.  This shows that the presence of
rich (Abell) clusters is closely related to the properties of
superclusters themselves.

Figs.~\ref{fig:5} and \ref{fig:7} show that the Northern slice of the
Sloan survey contains many very luminous DF-clusters, and in the Southern
slice the most luminous clusters are less luminous.  The number of
rich superclusters in the Northern slice of the SDSS is also much
higher than in the Southern slice.  The distribution of Abell clusters
on the celestial sphere is presented in Fig.~\ref{fig:11}.  The regions of
the SDSS EDR are marked by solid strips.  The Northern slice lies in a
region of space containing many rich superclusters.  The $\pm
10^{\circ}$ zone around the celestial equator in the right ascension
interval of the Northern slice contains 25 Abell superclusters in the
catalogue by Einasto et al. (2001), among these, there are 5 very rich
superclusters containing 7 or more Abell clusters.  The $\pm
10^{\circ}$ zone around the Southern slice has 12 Abell superclusters,
among which there is only one very rich system.

The discovery of large differences between the number and luminosity
of clusters in different slices of surveys is one of main results of
this study. We see that the number and richness of superclusters also
varies in a similar manner.  Evidently properties of clusters as well
as of superclusters are influenced by density perturbations of very large
scale.  

\section{Conclusions}

\begin{itemize}

\item{}  Galaxy systems have a similar shape in superclusters and voids: 
the dominant structural elements are single or multi-branching
filaments.  Massive superclusters have dominantly a multi-branching
morphology; less massive superclusters have various morphologies,
including compact, filamentary, multi-branching, and diffuse systems.

\item{} Selection effects distort the distribution of clusters and
  superclusters in two different ways. If a cluster has at least one
  galaxy bright enough to be included into the redshift survey, then
  the total luminosity of the cluster can be restored supposing that
  the galaxy luminosity function can be applied for individual
  clusters. If the cluster has no bright galaxies, it is lost. A
  respective correction can be made only at supercluster scales. 

\item{} There exists a strong dependence of cluster properties on the
density of the large-scale environment: clusters located in
high-density environments are a factor of $5 \pm 2$ more luminous than
clusters in low-density environments, and in extreme environmental density
regions the difference is even larger.

\item{} There exists a large difference between the properties of clusters
and superclusters in the Northern and Southern slices of the SDSS EDR
survey: the clusters and superclusters in the Northern slice are more
luminous than those in the Southern slice by a factor of 2.  This
difference may be due to differences in the location of slices with
respect to the very large-scale environment.

\end{itemize}

\section*{Acknowledgements}

I thank Maret Einasto, Gert H\"utsi, Enn Saar, Douglas Tucker, Volker
M\"uller, Pekka Hein\"am\"aki, Heinz Andernach and Erik Tago for
collaboration and for the permission to use our common results in this
review talk.  The present study was supported by the Estonian Science
Foundation grant ETF 4695, and by grant TO 0060058S98.  I thank Fermilab
and the Astrophysikalisches Institut Potsdam (DFG-grant 436 EST
17/2/01) where part of this study was performed, for hospitality.

\end{document}